# Joint MR sequence optimization beats pure neural network approaches for spin-echo MRI super-resolution

Hoai Nam Dang[1], Vladimir Golkov[2,3], Thomas Wimmer[2,3], Daniel Cremers[2,3], Andreas Maier[1]
Moritz Zaiss[1]

[1]Friedrich-Alexander-Universität Erlangen-Nürnberg (FAU)
[2]Technical University of Munich
[3]Munich Center for Machine Learning

HoaiNam.Dang@uk-erlangen.de

**Abstract.** Current MRI super-resolution (SR) methods only use existing contrasts acquired from typical clinical sequences as input for the neural network (NN). In turbo spin echo sequences (TSE) the sequence parameters can have a strong influence on the actual resolution of the acquired image and have consequently a considerable impact on the performance of the NN. We propose a known-operator learning approach to perform an end-to-end optimization of MR sequence and neural network parameters for SR-TSE. This MR-physics-informed training procedure jointly optimizes the radiofrequency pulse train of a proton density- (PD-) and T2-weighted TSE and a subsequently applied convolutional neural network to predict the corresponding PDw and T2w super-resolution TSE images. The found radiofrequency pulse train designs generate an optimal signal for the NN to perform the SR task. Our method generalizes from the simulation-based optimization to in vivo measurements and the acquired physics-informed SR images show higher correlation with a time-consuming segmented high-resolution TSE sequence compared to a pure network training approach.

**Keywords:** super-resolution, turbo spin echo, joint optimization

## 1    Introduction

Magnetic resonance imaging plays an essential role in clinical diagnosis by acquiring the structural information of biological tissue. Spatial resolution is a crucial aspect in MRI for the precise evaluation of the acquired images. However, there is an inherent trade-off between the spatial resolution of the images and the time required for acquiring them [1]. In order to obtain high-resolution (HR) MR images, patients are required to remain stable in the MR scanner for long time, which leads to patients' discomfort and inevitably introduces motion artifacts that again compromise image quality and actual resolution [2]. Since super-resolution (SR) can improve the image quality without changing the MRI hardware, this post-processing tool has been widely used to overcome the challenge of obtaining HR MRI scans [3]. Using model-based methods like interpolation algorithms [4] and iterative deblurring algorithms [5] or learning-based methods such as dictionary learning [6], SR achieved the restoration of fine structures and contours. In recent years, deep learning has become a main-stream approach for super-resolution imaging, and a number of neural network-based SR models were proposed [7]. Among the proposed model- or learning-based methods, convolutional neural networks (CNN) produce superior SR results with better clarity and less artifacts [8]. Super-resolution for MRI data has been only recently applied [9-12]. In [10] a CNN is proposed for cardiac MRI to estimate an end-to-end non-linear mapping between the upscaled low-resolution (LR) images and corresponding HR images to rebuild a HR 3D volume. In other work, motion compensation for the fetal brain was achieved by a CNN architecture [11] to solve the 3D re-construction problems. SR MRI has been also applied to low-field MR brain imaging [12]. However, these existing methods only used single contrast MRI images and did not make full use of multi-contrast information. In the clinical routine, T1, T2 and PD weighted images are often acquired together for diagnosis with complementary information. Although each weighted image highlights only certain types of tissues, they reflect the same anatomy, and can provide synergy when used in combination [8].

Fast imaging techniques like Turbo-Spin-Echo (TSE) [13] sequences can also be utilized to sample more data in given timeframe, thus allowing a higher resolution. However, due to the long echo-train duration the T2-decay is significant during the signal acquisition. This process acts as a voxel-T2-dependent k-space filter that lowers the actual resolution w.r.t the nominal resolution due to a broadening of the point-spread-function (PSF) [14]. However, by adjusting the refocusing radiofrequency (RF) pulses, the signal decay can be reduced during the TSE echo-train [15]. The RF pulse train strongly influence the signal dynamic in a highly complex fashion, as each RF pulse affects all future signal contributions. Current MRI super-resolution methods use contrasts acquired from typical clinical protocols as input for the neural network and disregard the influence of the MR sequence parameters for optimization. Using so-called known operator learning [16], we propose an approach that utilizes a MR physics model during the optimization to not only train a neural network for super-resolution, but also adapt the refocusing RF pulses to directly influence the PSF. This approach also allows the use of the uncorrupted theoretical contrast as ground truth, which is only available during the simulation. By using two different encoding schemes in our sequences, we gain additional information from the two different contrasts PD and T2 that are used as input for the CNN and both will have different PSFs, thus provide valuable information for the SR task. Both sequences are optimized jointly to allow generation of optimal contrasts for the SR task of the neural network.

The main contribution and the novelty of our work is the end-to-end optimization of MR sequence and neural network parameters for super-resolution TSE. For this purpose, we use a fully differentiable Bloch simulation embedded in the forward propagation to jointly optimize the RF pulse train of proton density (PD) and T2 weighted TSE sequences and a subsequent applied convolutional neural network to predict the corresponding PDw and T2w super-resolution TSE images. The ground truth targets are directly generated by the simulation and represent the uncorrupted MR contrast. Our jointly optimized approach is compared to a network trained on a TSE with a 180° RF pulse train. The optimized sequences and networks are verified at the real scanner system by performing in vivo measurements of a healthy subject and compared to a highly segmented, high-resolution vendor-provided sequence.

## 2    Theoretical Background

2.1    Image Degradation in TSE-Sequences.

When there is no relaxation decay during the echo-train, the k-space signal obtained from a TSE pulse sequence $S(k_x, k_y)$ yields the true spatial distribution of the theoretical transverse magnetization $M_\perp(x, y)$ via the Fourier transform (FT):

$$M_\perp(x,y) = \int \int_{k_x,k_y} S(k_x, k_y) \cdot e^{i(k_x x + k_y y)} dk_x dk_y \qquad (1)$$

When considering the signal relaxation behavior during acquisition an additional filtering function in k-space, the Modulation Transfer Function (MTF) for each tissue type (tt), has to be applied:

$$\begin{aligned}
\widetilde{M}_\perp(x,y) &= \sum_{tt} M(x,y) \\
&= \sum_{tt} \int_{k_x,k_y} S_{tt}(k_x, k_y) \cdot MTF_{tt}(k_x, k_y) \cdot e^{i(k_x x + k_y y)} dk_x dk_y \\
&= \sum_{tt} \int_{k_x,k_y} S_{tt}(k_x, k_y) \cdot e^{i(k_x x + k_y y)} dk_x dk_y \\
&\quad * \int_{k_x,k_y} MTF_{tt}(k_x, k_y) \cdot e^{i(k_x x + k_y y)} dk_x dk_y \\
&= \sum_{tt} M_{\perp,tt}(x,y) \cdot B_{tt}(x,y),
\end{aligned}$$

where $*$ denotes a convolution and $B(x, y)$ is a blur kernel equal to the Fourier-transformed MTF. For a single-shot 180° TSE sequence with constant RF pulses the filtering function in k-space can be described as:

$$MTF_{tt}(k_x, k_y) = \sum_{tt} e^{-\frac{t(k_x,k_y)}{T_{2tt}}}, \qquad (2)$$

which is unique for each tissue type with different T2 value. Using variable RF pulses, the MTF can become more homogeneous across the k-space and therefore reduce the width of the PSF.

## 3 Methods

### 3.1 Sequences

A single-shot 2D TSE sequence is being used as default sequence for our optimization. The acquisition time for the single-shot 2D TSE is 0.76 s at 1.56 mm in-plane resolution. Single slice acquisition was used for all sequences. The refocusing RF pulses of the PDw TSE with TE=12 ms and T2w TSE with TE=96 ms were optimized jointly. The PDw TSE sequence uses a centric phase-encoding reordering. For T2w imaging the centric phase-encoding reordering is shifted to have the central k-space line encoded at the given echo time TE, which for TE=96 ms is at the 8th echo. Other parameters were as follows: acquisition matrix of 128×128, undersampling factor in phase: 2x, reconstructed with GRAPPA[17], FOV=200 mm×200 mm, slice thickness of 8 mm and bandwidth of 133 Hz/pixel. For all sequences, the 90° excitation pulse was kept fixed.

### 3.2 Simulation & Optimization

All simulations and optimizations were performed in a fully differentiable Bloch simulation framework [18]. The framework generates MR sequences and corresponding reconstruction automatically based on the target contrast of interest. The optimization is carried out in an MR scanner simulation environment mirroring the acquisition of a real MR scanner. The forward simulation consists of a chain of tensor-tensor multiplication operations, representing the Bloch equations, that are differentiable in all parameters and supports an analytic derivative-driven nonlinear optimization. The entire process – MRI sequence, reconstruction, and evaluation – is modelled as one computational chain and is part of the forward and backwards propagation during the optimization, as depicted in Figure 1. The optimization problem is described by:

$$\Psi^*, \Theta^* = \underset{\Psi, \Theta}{\mathrm{argmin}} \left( \sum_i \left\| M_{\perp, i} - \mathrm{NN}_\Theta \left( RECO(\mathrm{SCAN}_\Psi(P_i)) \right) \right\|_p \right), \tag{3}$$

where $\Psi$ are the optimized sequence parameters and $\Theta$ the neural network parameters. For given tissue maps $P_i$ for each Voxel $i$ the Bloch simulation $SCAN$ outputs the MR signal and is reconstructed by the algorithm $RECO$. Signal simulations were performed by a fully differentiable extension of the Extended Phase Graph (EPG) [19] formalism. The simulation is done with PyTorch [20] complex-valued datatype and outputs a complex-valued signal. The forward simulation outputs the TSE signal which is conventionally reconstructed to magnitude images, and in addition the corresponding contrast as ground truth target as given in equation [1]. For the SR network a CNN, DenseNet [21] was adapted, which receives the magnitude TSE images of PDw and T2w TSE as input. To prevent scaling discrepancy, the TSE images are normalized to have maximum value of 1 before applying the CNN for both cases in simulation and in vivo. The DenseNet consist of 4 Dense blocks (Convolution->BatchNorm->PReLu->Concat) followed by an UpsampleBlock (bicubic upsampling->Convolution->BatchNorm->PReLu) and a final CNN Layer. Each convolution had a 3×3 kernel size, except for the first layer with a kernel size of 7×7. In total, the model had 174,706 trainable parameters. Of the sequence parameters Ψ, the amplitude of the refocusing RF pulses of the TSE sequences and the NN parameters Θ were optimized jointly in an end-to-end training procedure using the Adam optimizer [22] in PyTorch. We follow a known-operator approach [16], where the conventional reconstruction, including parallel imaging by means of GRAPPA, is fixed, but fully differentiable. The simulation is fully differentiable and all parameters except the refocusing RF pulses are fixed. The gradient update propagates back through the whole chain of differentiable operators. The complete RF pulse train and CNN are updated at each iteration step. The RF pulses are initialized with random values around 50° and standard deviation of 0.5°.

The training data consisted of synthetic brain samples based on the BrainWeb [23] database. The fuzzy model segments were filled with in vivo-like tissue parameters: Proton density PD values were taken from [24], T1 and T2 from [25], T2' was calculated from T2 and T2* values [26] and diffusion coefficient D was taken from [27]. B0 and B1 were assumed to be without inhomogeneities. In total, 19 subject volumes each consisting of 70 slices were used as training data and one separate subject volume as test dataset. The simulation uses coil sensitivity maps acquired at the MR system and calculated using ESPIRiT [28]. The optimizations were performed on an Intel Xeon E5-2650L with 256GB RAM. A full optimization on CPU took 4 days with memory consumption of 230GB RAM. The learning rate for the model parameters and sequence parameters were lr_model=0.001, lr_rf=0.01, respectively. Other hyperparameters of the optimization were: batch size = 1, n_epoch=10, damping factors of Adam (0.9, 0.999).

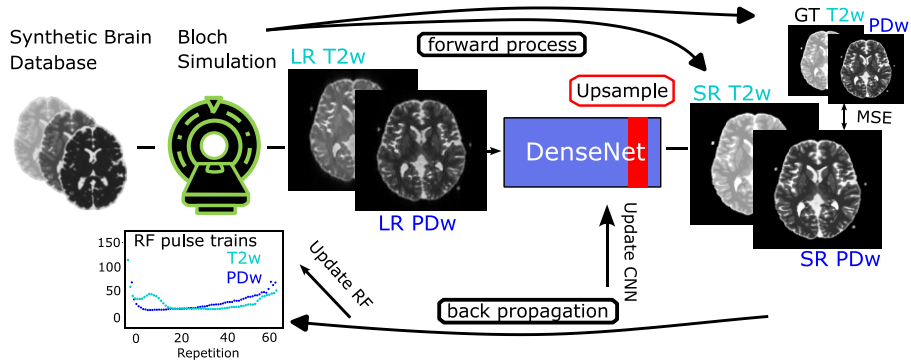

**Fig. 1.** Overview of the proposed processing pipeline: The MR signal of PDw and T2w TSE is simulated for a given RF pulse train; GRAPPA reconstruction and SR CNN are applied subsequently. The output is compared to the actual theoretical uncorrupted HR contrasts at TE$_{eff}$ and a gradient descent is performed to update refocusing FA and NN parameters, simultaneously.

### 3.3 Data acquisition at a real MR system

After the optimization process, all sequences were exported using the Pulseq standard [29] and the pypulseq tool [30]. Pulseq files could then be interpreted on a real MRI scanner including all necessary safety checks, and were executed on a PRISMA 3T scanner (Siemens Healthineers, Erlangen, Germany) using a 20-channel head coil. Raw data were automatically sent back to the terminal and the same reconstruction pipeline used for the simulated data was used for measured images. As high-resolution reference a vendor-provided TSE sequence was acquired with following parameters: 32-shot segmented, GRAPPA2, TE=12/96 ms, TR=12 s, FOV=200 mm×200 mm, matrix of 256×256, FA=180°. All MRI scans were under approval of the local ethics board and were performed after written informed consent was obtained. Measurements were performed on a healthy volunteer.

### 3.4  Reconstruction and Evaluation

Signals of the TSE sequences were reordered, and reconstructed with GRAPPA. The optimization was based solely on magnitude images. Structural similarity index measure (SSIM) [31] and the Peak Signal-to-Noise Ratio (PSNR) were calculated for evaluation of simulation and in vivo measurements w.r.t. the simulated ground truth and the HR segmented in vivo measurement, respectively. The evaluation was performed in Matlab [32] with the build-in functions for SSIM and PSNR.

## 4  Results

### 4.1  Qualitative Visual Results

The original LR TSE image with the zero-filled image and the reconstructed SR image are compared to our optimized RF pulse train design and a conventional 180° RF pulse train TSE sequence for each contrast in Figure 2. The optimization process can be seen in Supporting Figure S1 and Supporting Animation S3. Starting from the initialized values, the RF pulses converge to the optimal RF pulse train, while the NN parameters are optimized simultaneously. The converged RF pulse state has been found to be independent from the initialization. The final optimized RF pulse design for the PDw and T2w TSE sequences is shown in Figure 2a. It can be observed, that in all cases the SR image leads to an improvement over the LR TSE image by showing clearer resolved borders between white and gray matter. The optimized RF pulse train further improves the nominal resolution, which can be observed by a clear increase of sharpness of the sulcus between Gyrus cinguli and Gyrus frontalis superior as indicated by the red arrows. The optimized sequence and CNN translate well to in vivo measurements, where similar improvements as seen in the simulated images can be observed (Figure 2d,e).

### 4.2  Quantitative Metrics Results

Table 1 and Table 2 report the quantitative metrics scores of PSNR and SSIM for the images shown in Figure 2. The quantitative metrics agree with our visual observations and show that our end-to-end optimization approach performs better than the SR based on existing conventional 180° TSE sequence data only. Compared to the acquisition time of the segmented reference sequence with 192.85s, our optimized single-shot sequence only requires an acquisition time of 0.76s. Thus, the SR performance can potentially be further increased by a multi-shot scan sacrificing a little more time. To find the best procedure multiple ablation studies were performed (Supporting Information).

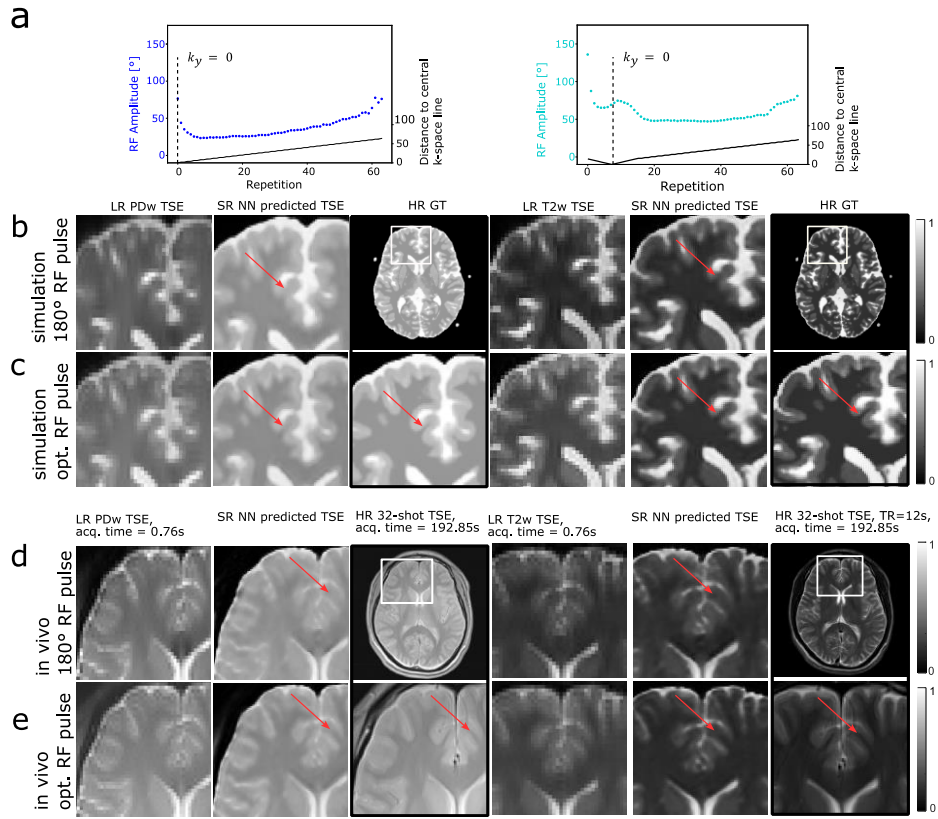

**Fig. 2.** (a) Optimized RF pulse train and phase encoding for both contrast (due to centric reordering, the central k-space line k_y=0 is acquired at repetition 0 and 7 for the PDw and T2w TSE, respectively). (b) Simulation of a static 180° RF pulse train and (c) optimized RF pulse train compared to the uncorrupted ground truth. (d) In vivo measurements of a static 180° RF pulse train and (e) optimized RF pulse train compared to the vendor's TSE sequence is shown as a high-resolution reference. In both cases the improvement of the optimized RF pulses over the constant 180° RF pulse train can be observed by a better resolved border between white and gray matter as indicated by the red arrows. SSIM and PSNR values are shown in Table 1 and 2.

**Table 1.** PSNR and SSIM for simulation in Figure 2 using the high-resolution GT as reference.

|  | PSNR - PDw | SSIM - PDw | PSNR – T2w | SSIM – T2w |
| --- | --- | --- | --- | --- |
| LR – 180° | 10.4 | 0.39 | 11.6 | 0.40 |
| ZF – 180° | 14.7 | 0.73 | 22.2 | 0.79 |
| HR NN – 180° | 28.7 | 0.93 | 29.4 | 0.93 |
| LR – opt. FA | 10.8 | 0.38 | 11.2 | 0.39 |
| ZF – opt. FA | 17.5 | 0.75 | 23.6 | 0.80 |
| HR NN – opt. FA | **30.6** | **0.97** | **31.1** | **0.95** |

**Table 2.** PSNR and SSIM for in vivo measurements in Figure 2 using the segmented high-resolution TSE as reference.

| | PSNR - PDw | SSIM - PDw | PSNR – T2w | SSIM – T2w |
|---|---|---|---|---|
| LR – 180° | 12.6 | 0.35 | 16.2 | 0.38 |
| ZF – 180° | 14.1 | 0.59 | 22.5 | 0.70 |
| HR NN – 180° | 23.2 | 0.84 | 26.2 | 0.89 |
| LR – opt. FA | 13.2 | 0.37 | 15.9 | 0.38 |
| ZF – opt. FA | 15.5 | 0.65 | 22.8 | 0.70 |
| HR NN – opt. FA | **26.0** | **0.87** | **27.0** | **0.90** |

## 5      Discussion

We demonstrated a new end-to-end learning process for TSE super-resolution by jointly optimizing refocusing RF pulse trains and neural network parameters. This approach utilizes a differentiable MR physics simulation embedded in the forward and backward propagation. The joint-optimization outperforms a pure neural network training. Although our approach is solely based on simulated data, the optimized sequence and trained CNN translate well to in vivo data. By using simulation-based training data, we are able to use the theoretical uncorrupted contrast as ground truth target. Apart from the expensive acquisition of HR in vivo data, real measured target data also have inherent drawbacks compared to their LR counterpart. Due to the longer scan time motion artifacts become more significant and to acquire the same contrast the bandwidth has to be increased, leading to a decrease of SNR [33]. However, we also admit the limitation of a simulation-based optimization, as the performance is bound to the accuracy of the model behind the simulation. We can observe in our results, that our approach is not able to resolve small vessel structures, as these are not existing in our synthetic brain database. Fortunately, the NN does not hallucinate details, when encountering these structures. Using real measured data to finetune the trained network could be a possible solution for this problem. Another way could be including uncertainty quantification layers [34] in the CNN to handle unknown structures. Our approach is compatible with any network architecture e.g. [35-37] to further improve the SR task. Furthermore, the training objective can be also extended to requirements on the MR sequence by including constraints in the loss function e.g. reduced RF pulse amplitudes for decrease of energy deposition SAR or increase of SNR.

   To conclude, we propose an end-to-end optimization of MR sequence and neural network parameters for TSE super-resolution. This flexible and general end-to-end approach benefits from a MR physics informed training procedure, allowing a simple target-based problem formulation, and outperforms pure neural network training.

**References**

1. Plenge, E. et al.: Super-resolution methods in MRI: can they improve the trade-off between resolution, signal-to-noise ratio, and acquisition time? Magn Reson Med. 68, 1983–1993 (2012).
2. Afacan, O. et al.: Evaluation of motion and its effect on brain magnetic resonance image quality in children. Pediatr Radiol. 46, 1728–1735 (2016).
3. Reeth, E. Van, Tham, I.W.K., Heng Tan, C., Loo Poh, C.: Super-Resolution in Magnetic Resonance Imaging: A Review. (2012).
4. Keys, R.G.: Cubic Convolution Interpolation for Digital Image Processing. IEEE Trans Acoust. 29, 1153–1160 (1981).
5. Hardie, R.: A Fast Image Super-Resolution Algorithm Using an Adaptive Wiener Filter. IEEE Transactions on Image Processing. 16, 2953–2964 (2007).
6. Yang, J., Wright, J., Huang, T.S., Ma, Y.: Image super-resolution via sparse representation. IEEE Transactions on Image Processing. 19, 2861–2873 (2010).
7. Bashir, S.M.A., Wang, Y., Khan, M., Niu, Y.: A Comprehensive Review of Deep Learning-based Single Image Super-resolution. PeerJ Comput Sci. 7, 1–56 (2021).
8. Lyu, Q. et al.: Multi-Contrast Super-Resolution MRI through a Progressive Network. IEEE Trans Med Imaging. 39, 2738–2749 (2020).
9. Kaur, P., Sao, A.K., Ahuja, C.K.: Super Resolution of Magnetic Resonance Images. J Imaging. 7, (2021).
10. Oktay, O. et al.: Multi-input cardiac image super-resolution using convolutional neural networks. Lecture Notes in Computer Science (including subseries Lecture Notes in Artificial Intelligence and Lecture Notes in Bioinformatics). 9902 LNCS, 246–254 (2016).
11. Song, L. et al.: Deep robust residual network for super-resolution of 2D fetal brain MRI. Scientific Reports 2022 12:1. 12, 1–8 (2022).
12. de Leeuw den Bouter, M.L., Ippolito, G., O'Reilly, T.P.A., Remis, R.F., van Gijzen, M.B., Webb, A.G.: Deep learning-based single image super-resolution for low-field MR brain images. Scientific Reports 2022 12:1. 12, 1–10 (2022).
13. Hennig, J., Nauerth, A., Friedburg, H.: RARE imaging: A fast imaging method for clinical MR. Magn Reson Med. 3, 823–833 (1986).
14. Qin, Q.: Point Spread Functions of the T2 Decay in k-Space Trajectories with Long Echo Train. Magn Reson Imaging. 30, 1134 (2012).
15. Busse, R.F., Hariharan, H., Vu, A., Brittain, J.H.: Fast spin echo sequences with very long echo trains: Design of variable refocusing flip angle schedules and generation of clinical T2 contrast. Magn Reson Med. 55, 1030–1037 (2006).
16. Maier, A.K. et al.: Learning with known operators reduces maximum error bounds. Nature Machine Intelligence 2019 1:8. 1, 373–380 (2019).
17. Griswold, M.A. et al.: Generalized autocalibrating partially parallel acquisitions (GRAPPA). Magn Reson Med. 47, 1202–1210 (2002).
18. Loktyushin, A., et al. "MRzero-Automated discovery of MRI sequences using supervised learning." Magn Reson Med 86.2 (2021): 709-724.
19. Hennig, J., Weigel, M., Scheffler, K.: Calculation of Flip Angles for Echo Trains with Predefined Amplitudes with the Extended Phase Graph (EPG)-Algorithm: Principles and Applications to Hyperecho and TRAPS Sequences. Magn Reson Med. 51, 68–80 (2004).
20. Paszke, A. et al.: PyTorch: An Imperative Style, High-Performance Deep Learning Library. Adv Neural Inf Process Syst. 32, (2019).

21. Huang, G., Liu, Z., Van Der Maaten, L., Weinberger, K.Q.: Densely Connected Convolutional Networks. Proceedings - 30th IEEE Conference on Computer Vision and Pattern Recognition, CVPR 2017. 2017-January, 2261–2269 (2016).
22. Kingma, D.P., Ba, J.L.: Adam: A Method for Stochastic Optimization. 3rd International Conference on Learning Representations, ICLR 2015 - Conference Track Proceedings. (2014).
23. Cocosco, C., Kollokian, V., Kwan, R., Evans, A.C.: BrainWeb: Online Interface to a 3D MRI Simulated Brain Database. undefined. (1997).
24. Bojorquez, J.Z., Bricq, S., Acquitter, C., Brunotte, F., Walker, P.M., Lalande, A.: What are normal relaxation times of tissues at 3 T? Magn Reson Imaging. 35, 69–80 (2017).
25. Peters, A.M. et al.: T2* measurements in human brain at 1.5, 3 and 7 T. Magn Reson Imaging. 25, 748–753 (2007).
26. Le Bihan, D. et al.: Diffusion tensor imaging: Concepts and applications. Journal of Magnetic Resonance Imaging. 13, 534–546 (2001).
27. Lim, Y., Bliesener, Y., Narayanan, S., Nayak, K.S., Yongwan Lim, C., Hsieh, M.: Deblurring for spiral real-time MRI using convolutional neural networks. (2020).
28. Sandino, C.M., Lai, P., Vasanawala, S.S., Cheng, J.Y., Christopher Sandino, C.M.: Accelerating cardiac cine MRI using a deep learning-based ESPIRiT reconstruction for Magnetic Resonance in Medicine. (2020).
29. Layton, K.J. et al.: Pulseq: A rapid and hardware-independent pulse sequence prototyping framework. Magn Reson Med. 77, 1544–1552 (2017).
30. Ravi, Keerthi, Sairam Geethanath, and John Vaughan. "PyPulseq: A Python Package for MRI Pulse Sequence Design." Journal of Open Source Software 4.42 (2019): 1725.
31. Wang, R., Tao, D.: Recent Progress in Image Deblurring. (2014).
32. The MathWorks, Inc. (2022). MATLAB version: 9.9.0 (R2020b. Accessed: March 09, 2023. Available: https://www.mathworks.com
33. Portnoy, S., Kale, S.C., Feintuch, A., Tardif, C., Pike, G.B., Henkelman, R.M.: Information content of SNR/resolution trade-offs in three-dimensional magnetic resonance imaging. Med Phys. 36, 1442–1451 (2009).
34. Abdar, M. et al.: A review of uncertainty quantification in deep learning: Techniques, applications and challenges. Information Fusion. 76, 243–297 (2021).
35. Li, G., Lyu, J., Wang, C., Dou, Q., Qin, J.: WavTrans: Synergizing Wavelet and Cross-Attention Transformer for Multi-contrast MRI Super-Resolution. Lecture Notes in Computer Science (including subseries Lecture Notes in Artificial Intelligence and Lecture Notes in Bioinformatics). 13436 LNCS, 463–473 (2022).
36. Sui, Y., Afacan, O., Gholipour, A., Warfield, S.K.: MRI Super-Resolution Through Generative Degradation Learning. Lecture Notes in Computer Science (including subseries Lecture Notes in Artificial Intelligence and Lecture Notes in Bioinformatics). 12906 LNCS, 430–440 (2021).
37. Lyu, Q., Shan, H., Wang, G.: MRI Super-Resolution with Ensemble Learning and Complementary Priors. IEEE Trans Comput Imaging. 6, 615–624 (2019).

# Supplementary Material

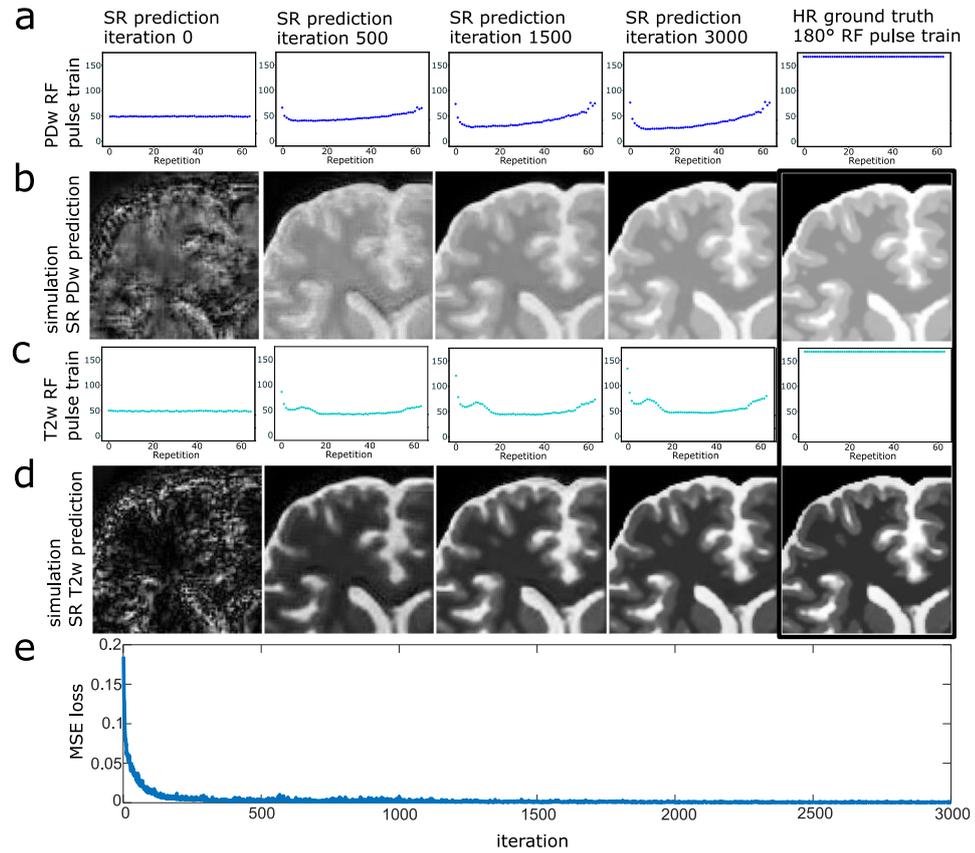

**Fig. S1.** Optimization process at different iterations. The optimized RF pulse trains for the PDw TSE in (a) and T2w TSE in (c) are shown for different iterations. The corresponding SR prediction for PDw in (b) and T2w in (d) are compared to the HR ground truth contrast in the last column. Joint MSE loss is shown in (e) over the iterations.

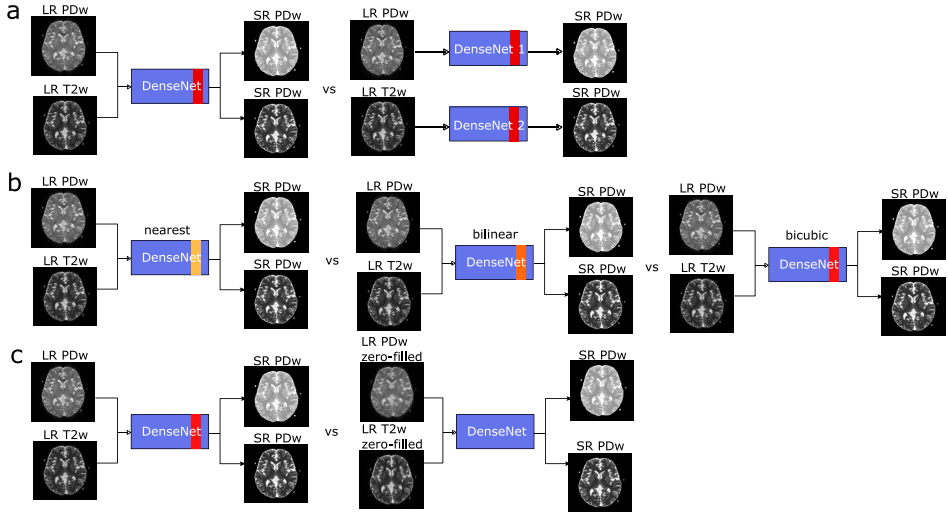

**Fig. S2.** Schematic view of the ablation studies. Only the neural network was trained in these studies with a fixed RF pulse train. We performed systematically three different ablation studies and each study was evaluated by using the quantitative metrics reported in Table 3. (a) We investigated the performance of two separate networks with single contrast input compared to a network using two contrast as input. The multi-contrast approach demonstrates higher SSIM and PSNR scores compared to the single-contrast approach. (b) Different upsampling methods were compared against each other. Between nearest neighbor, bilinear and bicubic interpolation a slight improvement by using bicubic interpolation could be observed. (c) We investigated the performance between using the zero-filled image as input instead of the original LR image and the upsampling layer in the CNN. Using LR image as input with an upsampling layer in CNN demonstrated better performance compared to the zero-filled image input.

**Table S3.** Metrics PSNR and SSIM for the ablation studies. Average over both contrasts are shown for the metric scores.

|  | single-contrast | Multi-contrast | nearest | bi-linear | bicubic | zero-filled input | LR input & up-sampling |
|---|---|---|---|---|---|---|---|
| PSNR | 29.8 | **30.9** | 30.6 | 30.8 | **30.9** | 29.4 | **30.9** |
| SSIM | 0.94 | **0.96** | 0.95 | 0.96 | **0.96** | 0.93 | **0.96** |

VideoS3.mp4 - https://streamable.com/yionx2

**Animation S3.** Example for a training procedure over the training set. PDw and T2w TSE images are simulated for the current RF pulse trains and used as input for the CNN, which predicts the SR images for both contrasts, respectively. The MSE loss is calculated between neural network output and the HR ground truth. For iteration 100, 500, 1500 and 2500 an in vivo measurement was acquired.